\documentclass{aa}

\usepackage{epsf}
\begin{document}

   \thesaurus{22     
              (03.09.1;  
               03.20.9;  
               13.07.1;  
               13.07.2)} 

   \title{The MAGIC Telescope - Prospects for GRB research}


   \author{D. Petry\inst{1} for the MAGIC Telescope Collaboration\thanks{
       The MAGIC Telescope Collaboration is presently formed by approx. 50 members of the
           following institutions: IFAE, Universitat Aut\`onoma de
            Barcelona, Spain / Crimean Astrophysical Observatory, Crimea, Ukraine /
            Universit\"{a}ts-Sternwarte, G\"{o}ttingen, Germany / Division of Experimental 
            Physics, University of Lodz, Poland / Facultad de Ciencias Fisicas, Universidad
            Complutense, Madrid, Spain / Institute for Nuclear Research, Moscow, Russia /
            Max-Planck-Institut f\"{u}r Physik, M\"{u}nchen, Germany /
            Dipartimento di Fisica, Universita' di Padova, Italy / Dipartimento di Fisica,
            Universita' di Siena, Italy / Space Research Unit, Potchefstroom University,
            South Africa / Fachbereich Physik, Universit\"{a}t Wuppertal, Germany /
            Cosmic Ray Division, Yerevan Physics Institute, Armenia /
            Ruder Bo\v{s}kovi\'c Institute, Zagreb, Croatia} 
          }

   \offprints{D. Petry}

   \institute{$^1$Institut de F\'{\i}sica d'Altes Energies, Universitat Aut\`onoma de
            Barcelona, 08193 Bellaterra, Spain\\
              email: petry@ifae.es
             }

   \date{Received ; accepted }

   \maketitle

   \begin{abstract}

  The Major Atmospheric Gamma-ray Imaging Cherenkov (MAGIC)
  Telescope collaboration is constructing a large Cherenkov telescope (17 m 
  diameter) for the exploration of the gamma-ray energy regime above
  10 GeV with high sensitivity. One of the highlights in the science
  program of this future observatory are the plans for fast follow-up
  observations of of GRBs. By ``fast'' we mean delays of less than 30 s
  between notification and the beginning of observations. The expected
  gamma counting rates are of the order of 100 Hz for a EGRET counting
  rate of 0.1 Hz above 100 MeV and a spectral index of 2.0 (this would
  correspond to a fluence of 10$^{-5}$ erg cm$^{-2}$ for 60 s burst 
  duration).  

  The good photon statistics will permit determination of spectra, search
  for cutoffs and measurement of light-curves with a time resolution of 
  the order of 1 s.

   \end{abstract}

%

\section{Introduction}
The Major Atmospheric Gamma-ray Imaging Cherenkov (MAGIC) Telescope 
will be a large imaging air Cherenkov telescope for ground-based
$\gamma$-ray observations above 10\,GeV.  Details of its design, the
scientific motivation, the feasibility of the project, and other issues
have been described elsewhere (Barrio et al. \cite{barrio}). 
The telescope is expected to become operational by the middle of 2001.

The telescope is optimised to achieve the lowest energy threshold and
highest flux sensitivity achievable with present technology. This makes it 
applicable to  a large range of astrophysical research fields:
\begin{itemize}
\item Blazars (study of EGRET blazars, possible discovery of additional sources) 
\item Cosmology (measurements of the near-infrared background via $\gamma$-$\gamma$
      absorption)
\item Investigation of the pulsed $\gamma$-ray emission from pulsars
\item Search for gamma-emission from Supernova remnants
      (and hence search for evidence of production and acceleration of cosmic rays)
\item Search for decay/annihilation line-emission from WIMPs clustering at the galactic centre
\item Identification of ``unidentified EGRET sources'' (position accuracy $\approx$ 1.2')   
\item Search for high-energy counterparts of gamma-ray bursts 
\end{itemize}
The last part of this scientific program will briefly be discussed in this
article.

\section{Fast follow-up observations}
In order to make fast follow-up observations of GRBs possible,
the MAGIC Telescope will have an interface (probably a socket connection) 
to the GRB Coordinate Network GCN (Barthelmy et al. \cite{barthelmy} and these proceedings) 
and a secondary  ground station for direct communication with HETE II 
(Ricker et al., these proceedings). HETE II is a dedicated GRB research satellite
which is expected to be launched in 2000. It
 will provide approx. 30 burst positions per year with accuracies
better than 10 arcmin. These notifications are expected to arrive with
a delay of less than 5\,s. 
The MAGIC Telescope is specially designed to have low inertia
such that the telescope can be positioned on any point in the sky
within less than 30\,s.  

In case of a notification, a fast check of the observability of the
GRB location will be performed. If the decision is positive, the
present observations will be stopped immediately and a special
fast drive will position the telescope on the GRB which will then be
observed for the remaining night-time and probably also the
following night in order to detect possible delayed emission.

Taking into account a decision time of 5 s at our site,
we expect the reaction time
between the actual start of the burst and the start of the follow-up
observation to be $30\pm10$ s. The MAGIC Telescope will thus be able
to perform observations of non-delayed emission of all bursts with 
durations above 30\,s. This is the
position of the rightmost peak in the BATSE burst duration ($T_{90}$) distribution 
(see e.g. Kouveliotou et al. \cite{kou}) and corresponds to a fraction
$\approx$ 33\,\% of all bursts which trigger BATSE on the 64 ms time scale.

The telescope will be built on the Canary Islands (Tenerife or La Palma).
Cherenkov telescopes can only observe during the night.
Observations will be possible up to zenith angles of $\approx 80^\circ$,
i.e. about 40 \% of the total sky ($4\pi$) will be accessible.
Assuming 30 \% of the nights to have bad weather and taking into account that
the presence of the moon can prevent observations of certain positions, 
we arrive at a duty cycle of $\approx$ 10 \%. 

The effective field of view of the photo-sensor camera of the MAGIC Telescope 
is 1.6$^\circ$ in diameter. We will therefore be able to safely observe any
of the positions provided by HETE II and also some from the GCN and expect
$\approx$ 5 serious immediate follow-up observations per year.  
For delayed emission (time-window of the order of an hour up to several days) this number
will be larger.  

\section{Expected performance}
\begin{figure}
\leavevmode
\centering
\epsfxsize=8.8cm
\epsffile{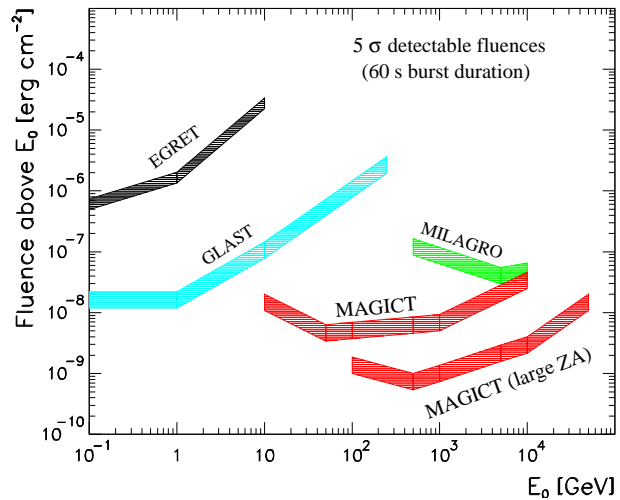}
\caption{\label{fig-magflusens} 
The fluence sensitivity of the MAGIC Telescope (MAGICT) compared to other experiments 
(MAGICT refers to observations at 
zenith angles up to $\approx$ 30$^\circ$, ``MAGICT large ZA'' to those at $\approx$ 
70$^\circ$). The fluence sensitivity at threshold $E_0$ is defined here as the total 
fluence above $E_0$ which is necessary for a $5 \sigma$ detection of the burst above
that threshold. A total burst duration of 60 s and a differential spectral index of 2.6
is assumed. For MAGICT the actual observation 
time is assumed to be only 30 s taking into account the reaction time. For all other detectors
the observation time is assumed to be equal to the burst duration.} 
\end{figure}

The expected performance in terms of sensitivity is summarised in 
Figure \ref{fig-magflusens}. This is the performance we expect for the second
phase of the project in which high quantum efficiency hybrid photo sensors will replace the
photo-multipliers of the telescope's camera. In the first phase our threshold
will be $\approx$ 30 GeV, in the second phase 10 GeV. It is not yet clear
how soon the second phase will follow the first. The sensitivity at energies
$>$ 30\,GeV is to a good approximation independent of these changes.

Given the diverse shape of GRB light-curves it is difficult to predict an average
counting rate for gamma-rays in successful burst observations. We note instead that the
MAGIC Telescope will have an effective collection area for primary gamma photons
of $10^8$\,cm$^2$ at the threshold rising to $10^9$\,cm$^2$ at 100 GeV. 
For a hypothetical counting rate of 0.1\,Hz for EGRET above 100 MeV, we expect 
(assuming a spectral index of 2.0) a counting rate of $\approx$ 100\,Hz above 10 GeV. 
A rate of $\approx$ 6 Hz for 30 s will suffice for a 5 $\sigma$ detection with
a moderate background rejection applied. For strong bursts such as GRB930131, 
gamma counting rates of the order of 1 kHz may occur. The data acquisition will
therefore be prepared to sustain such rates without additional dead-time. 

\section{Conclusions}
The MAGIC Telescope will be able to make a major contribution to GRB research
by providing high sensitivity measurements in the (for GRBs)
essentially unexplored regime above 10 GeV. The high counting rates in
strong bursts will also permit to study the shape of the light-curve in more
detail than previously possible (time resolutions of the order of 1 s). 
Search for small delays with respect to the low energy emission is thus possible.    
As an interesting side-result this may provide one of the best possible lower limits
to the quantum gravity energy scale as pointed out by Amelino-Camelia et al. 
(\cite{amelino}).

\end{document}